# Rendering-as-a-Service: Taxonomy and Comparison

Ruby Annette.J[a], Dr. Aisha Banu.W[b], Subash Chandran. P[a,b,*]

[a]Resaerch Scholar, BSAU, Chennai, India
[b]Professor, BSAU, Chennai, India

**Abstract**

The movies like the "Avathar" are a good example of the stunning visual effects that the animation could bring into a movie.

The 3D wireframe models are converted to 3D photorealistic images using a process called the rendering. This rendering process is offered as a service in the cloud, where the animation files to be rendered are split into frames and rendered in the cloud resources and are popularly known as Rendering-as-a-Service (RaaS). As this is gaining high popularity among the animators community, this work intends to enable the animators to: (a) Gain basic knowledge about Rendering-as-a-Service (RaaS). (b) Understand the variety in the RaaS service models through the taxonomy (c) Explore, compare and classify the RaaS services quickly using the tree-structured taxonomy of services. In this paper, the various characteristics of the RaaS services are organized in the form of a tree to enable quick classification and comparison of the RaaS services. To enhance the understandability, three popular RaaS services have been classified and verified according to the proposed tree-structured taxonomy.

## 1. Introduction

Evolution of distributed and parallel computing technologies has paved way for the concept of Render farms. An offline render farm usually contains a cluster of computers which are connected together in a network for rendering the animated models.[26, 27]. Each individual computer in the cluster is called a Render node. An animated scene usually contains many individual frames. In a render farm, each frame is rendered independently in different render nodes at the same time. This reduces the rendering time drastically. In order to distribute the tasks automatically to each render node, Rendering Job Management software (RJM) is used, which acts as a queue manager and assigns the rendering tasks to appropriate render nodes based on its scheduling policy[26]. The rendering task is offered as a service using the cloud computing technology as is known as Rendering-as-a-Service (RaaS).

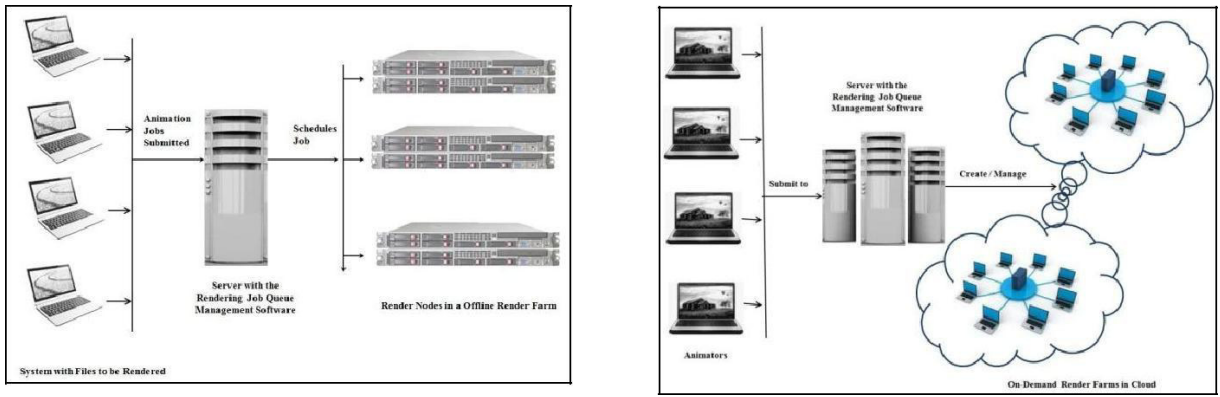

Fig. 1. (a) Offline Rendering ; (b) Cloud based Rendering.

The Rendering-as-a-Services is based on the principles of Cloud based rendering. The files to be rendered are uploaded and sent to the RaaS service providers and the rendering task is completed using the virtual machines in their render farms [26]. The figure 1a and 1b given above explains the offline and cloud rendering process. Using RaaS services on can determine the rendering deadline and increase the no of virtual machines accordingly to meet the deadline. The advantage of RaaS services is that, the user needs to pay only for the resources utilized on an hourly basis. Hence, the animation studios need not invest much in building their own render farms. However as the RaaS services follow different business models, the intended users of RaaS services like the animators, 3D studios, freelancers and students need to spend a lot of time to explore, analyse and identify the suitable RaaS service providers. In this work, we propose a tree structured taxonomy to classify the RaaS services effectively.

Many have explored the distributed computing technologies like the grid [1-5] and the cloud for the rendering purpose and had proved them to be fruitful [6-9, 16, 21, 22, 25]. There are many works focusing on the state of the art, research challenges, surveys and taxonomy of cloud services in general [6-15]. But very few on RaaS services like the work of Ruby et al [47] but none on the taxonomy of the RaaS services according to our literature survey. This paper is intended to bridge this gap. The contributions of this work are: a) To provide knowledge about the RaaS Services business models through the taxonomy as given in section 2. b) To enable the animators to compare the popular RaaS services easily and effectively using the tree structured taxonomy as illustrated with examples in the section 3. Finally, the Section 4 concludes the work with scope for further work.

## 2. Taxonomy design of RaaS services

The important characteristics of the RaaS services are defined and taxonomy has been designed. The taxonomy of RaaS services has a tree-structure. The Cloud renderfarm services also called as the RaaS is placed at the root of the tree. The other main characteristics of the service like the deployment models, delivery models, license type etc are defined in different levels of the tree nodes as given in figure 2a given below. The other characteristics specific to the IaaS and the PaaS levels are defined in a separate taxonomy tree to make the diagrams simple and easy to understand as given in figure 2b. The detailed taxonomy design and the levels in the design are explained in detail.

The taxonomy levels are:
| | | |
|---|---|---|
| L1. Delivery Models(DM) | L2a. Delivery Model Components(Comp) | L2.b Sub Components(SubComp) |
| L3. Deployment Models(DPM) | L4. License Type(LiT) | L5. User Group (UG) |
| L6. Formal Agreements(FA) | L7. Security Measures(SM) | L8. Payment Types (PT) |
| L9. Priced Components(PC) | L10: Pricing Models (PM) | |

The common IaaS and PaaS level Characteristics as given in figure 3 are:
I1. Operating System (OS)   I2. Development Tools(DT)   I3. Virtualization Technology used(VT)

The PaaS Levels in the taxonomy are:
P1.Cost Estimation Methods(CEM)　　P2. Rendering Job Manager (RJM)　　P3.RJMSoftwareLicenseTypes (JM_L)
P4. RJM Pricing Types (JM_P)　　P5. RJM Deployment Types (JM_D)　　P6. Rendering Software License (RSL)
P7. Render Engine Supported (RES)　P8. Software Supported(S/W)　　P9.Plugins supported (Plugin)
P10. Free Support Provided (FSup)　P11. Paid Support Offered (PSup)

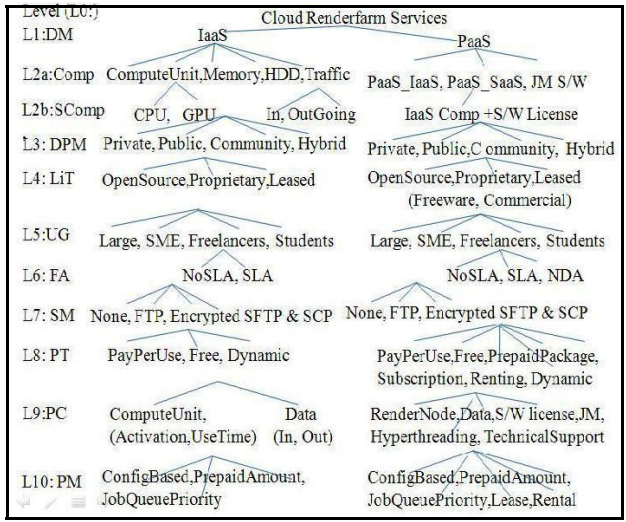
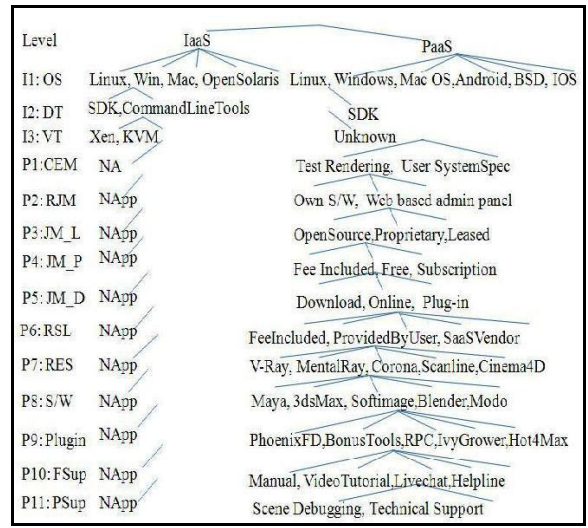

Fig. 2. (a) Main Taxonomy Levels ; (b) IaaS, PaaS Taxonomy Levels of RaaS services.

2.1 Main Characteristics of RaaS services

2.1.1 *RaaS Deployment Models*

The cloud renderfarm deployment models include public, private and hybrid renderfarms. In a public cloud renderfarm, the animation files to be rendered are split into individual frames and the job management software schedules the rendering job on the render nodes of the public render farm. The user is given the privilege of modify the settings like the lighting etc, pause or kill the rendering job or download the rendered files. In case of a Private cloud Renderfarm, the resources of a 3d studio are pooled up to speedup the rendering process. A hybrid cloud renderfarm is a combination of both the private and the public cloud renderfarms. Some of the popular Cloud render farm services are Amazon EC2[28], Rackspace[29], Renderingfox[30], Rebusfarm[31] etc. Community based Renderfarms are intended to enable pooling up of resources of the users in a community who share their resources for mutual benefit and are usually free. Examples include Vswarm[32].

2.1.2 *RaaS Delivery Models*

　　The three types of service models of cloud computing in general are: a) Infrastructure-as-a-Service (IaaS), Software-as-a-Service (SaaS) and Platform-as-a-Service (PaaS)[4]. However the cloud based renderfarms are delivered as IaaS or PaaS models. The SaaS is included in the PaaS model of services to provide a holistic rendering environment to the users. The IaaS delivery model enables the user to rent the required computing power like the compute unit, bandwidth, storage space etc on an hourly basis and paid only for the resources that had been used. Leading players include Amazon EC2[28], Rackspace[29]. Amazon EC2 supports Cloud based rendering through Axceleon's Cloud Fusion[33]. A detailed taxonomy of the IaaS services can be found in the work of Prodan et al[4]. PaaS offers a complete infrastructure to the animators which include the compute unit (PaaS_IaaS), software licenses of the software required like the rendering software (PaaS_ SaaS), the rendering job management tools etc. PaaS_ SaaS feature takes care of the license required for the software and plug-ins are used during the rendering process and the charges are usually included in the render node charges like in case of the Rebus farm[31], RevUp Render[36]. The animation related software vendors are also moving towards cloud services for example, Autodesk desktop subscription model (SaaS), lets the user to rent the software on the pay-as-u-go model[37] and Pixar[34] has introduced a Cloud based rendering tool called the RenderMan Pro Server [35] for Windows Azure.

*2.1.3. License type*

The software license type includes free, proprietary software or a leased license from a third party. The software licensing issues is one of the major problems in cloud computing[38-40]. Most of the cloud monitoring software and smaller cloud computing services are open-source based, as the small players often lack influence to push their proprietary software on the market effectively[39]. For example, the community based RaaS services are built on free and open source software platforms. In the IaaS type of services, the software license details are provided by the user if required. Whereas in the PaaS type of service, the PaaS_IaaS, PaaS_SaaS and the rendering job management software licensing issues are dealt by the RaaS Service provider and the user is usually charged a fee separately or may be included along with the cost of the render node. I some cases, the users need to provide the license details. Some of the services also arrange for the license through the SaaS vendors for an additional fee. For example, the services like the RenderSolve, RebusRenderFarm and the RenderCore provide render nodes with software license and the fee is included in the cost of the render node.

*2.1.4. Potential users of RaaS services*

Large animation studios have also turned towards the Cloud based Renderfarms to reduce the production cost and also satisfy the demand for high quality at the same time. For example "The Adventures of TinTin: the secret of the unicorn" used the Cloud based Renderfarms for rendering[41]. The other potential users of the RaaS services include the SME's (Small and Medium Enterprises), freelancers and the students. They often lack money to build their own render farms and turn to RaaS service providers to satisfy their needs of high rendering power on the pay as peruse model.

*2.1.5. Formal Agreements*

The formal agreement usually signed by the cloud service providers is the Service Level Agreement. However, the Non Disclosure Agreement (NDA) is also very important for the 3d animation studios. But most of the RaaS service providers of both the IaaS and PaaS type normally sign a Non Disclosure Agreement (NDA), which prevents copying of the animation model designs. But most of them have not given any information about the SLA except for a few like the Amazon EC2, and the details provided are limited.

*2.1.6 Free and paid technical support*

Most of the service Providers provide free technical support like User manuals, Video tutorials, live chat, 24/7 help line etc. while some other service providers also provide paid technical support for their customers like scene debugging etc.

*2.1.7 Cost*

The important criteria to be considered when considering the rendering cost of an animation file includes, the Pre-rendering cost estimation methods, payment models, priced components and the pricing models. The Pre-rendering cost estimation methods help the users to estimate the rendering cost of their files even before submitting them for actual rendering. The rendering cost is estimated through test rendering using per frame render time in user system. The payment models popular among the IaaS and PaaS services are the subscription, prepaid packages and monthly rental models and are usually priced based on various criteria like the job queue priority, configuration etc. The components usually priced are the render node unit which may be a CPU or a GPU and the data. The Activation time and the use time are generally charged. The other priced components include the storage space, incoming and outgoing traffic.

## 3. Classifying and Comparing RaaS services

In this section the characteristics of three popular RaaS services are compared. The detailed comparison chart is given as a table below in figure 3a. Using these details, the tree diagram of three services namely the Amazon EC2, RebusRenderfarm and VSwarm which belong to three different categories of IaaS, PaaS-Public and PaaS-Community are given in figure 4l and 4b.

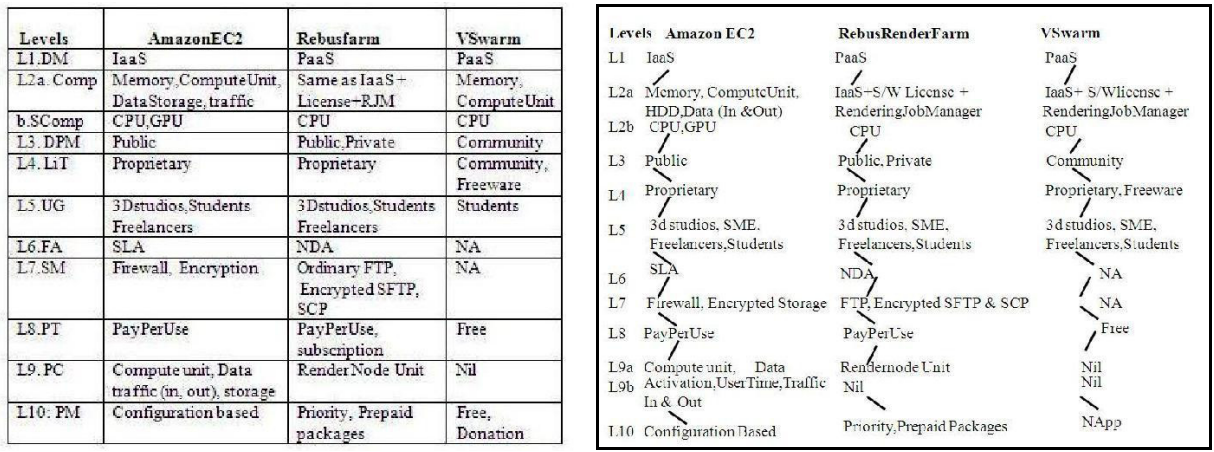

Fig. 3. (a) Main characteristics of RaaS services ; (b) Tree structured taxonomy of main characteristics of RaaS services.

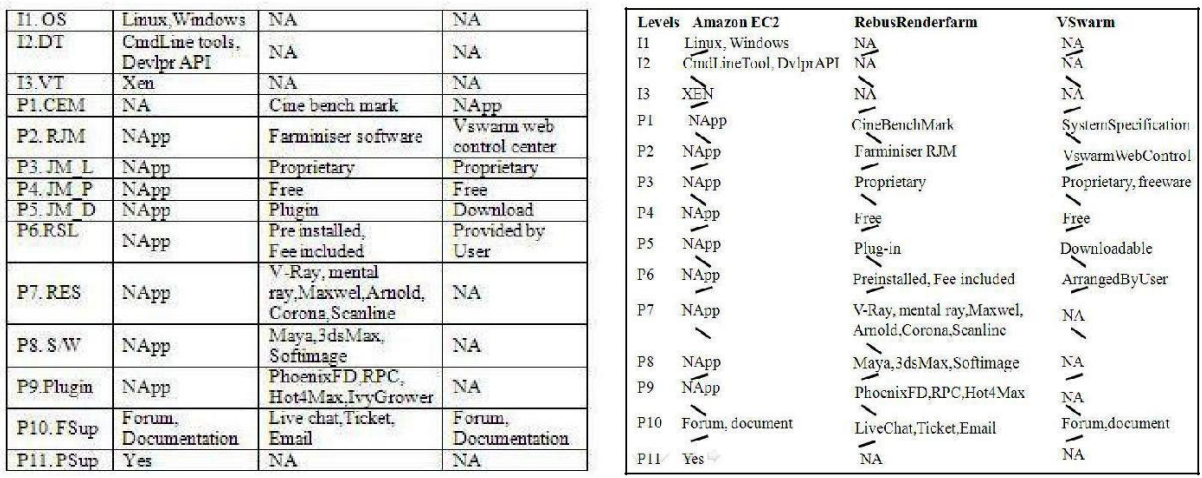

Fig. 4. (a) IaaS,PaaS characteristics of RaaS services ; (b) Tree structured taxonomy of IaaS,PaaS characteristics of RaaS services.

## 4. Conclusion and future work

This paper has explored, analysed and classified the RaaS services and a tree structured taxonomy of the RaaS services provides has been proposed. The tree structured taxonomy helps the users to identify the similarities and the differences among the interested services quickly and effectively. To enhance the understanding of the users like the animators and freelancers an illustrative example of the taxonomy of three popular services has been given in figure 3b and 4b. The taxonomy gives a clear idea to the animators about the various types of services being offered. They also gain insight about what to look for when choosing a service. As a future work, the taxonomy will be expanded and formed as an ontology that would facilitate the RaaS service selection in a more efficient manner. It is also clearly evident that the animators have to do a lot of research in selecting the suitable RaaS services and may need professional help to select, negotiate and monitor them. Thus in the future a Cloud Broker Service model for the selection, negotiation and monitoring of the RaaS service providers would be developed.